\documentclass{elsart}
\usepackage{hyperref}
\usepackage[cp1250]{inputenc}
\usepackage{harvard}
\usepackage{amsmath}
\usepackage{amssymb}
\usepackage{graphicx}
\usepackage{url}
\usepackage{booktabs}
\usepackage{array}
\begin{document}
\begin{frontmatter}
\title{Structural Changes on Warsaw's Stock Exchange: the end of Financial Crisis}
\author
{Paweł Fiedor} \ead{s801dok@wizard.uek.krakow.pl}
\address{Cracow University of Economics, Rakowicka 27, 31-510 Kraków, Poland}
%\maketitle

\begin{abstract}

In this paper we analyse the structure of Warsaw's stock market using complex systems methodology together with network science and information theory. We find minimal spanning trees for log returns on Warsaw's stock exchange for yearly times series between 2000 and 2013. For each stock in those trees we calculate its Markov centrality measure to estimate its importance in the network. We also estimate entropy rate for each of those time series using Lempel-Ziv algorithm based estimator to study the predictability of those price changes. The division of the studied stocks into 26 sectors allows us to study the changing structure of the Warsaw's stock market and conclude that the financial crisis sensu stricto has ended on Warsaw's stock market in 2012-13. We also comment on the history and the outlook of the Warsaw's market based on the log returns, their average, variability, entropy and the centrality of a stock in the dependency network.

\end{abstract}
\end{frontmatter}

\section{Introduction}

Financial markets are well defined complex systems. Nonetheless they do lack a fundamental theory behind their behaviour. Therefore they are studied not only by economists, but also by mathematicians and physicists. The lack of theory means that it's assumed the time series of stock returns are unpredictable \cite{Samuelson:1965}. Within this paradigm, time evolution of stock returns can only be described by random processes. The question then arises whether those random processes are uncorrelated for different stocks or whether in fact there are common economic factors driving the price formation processes for different stocks. Common economic factors were never reliably found, nonetheless tools and procedures developed to model physical systems \cite{Mandelbrot:1963,Kadanoff:1971,Mantegna:1991} can be used to characterise the interdependencies of different stocks. 

The recent financial crisis renders the investigation into the complex nature of the financial markets and their dynamical properties more important than ever. We are therefore looking at the Warsaw's market between 2000 and 2013 to see what changes in the structure and characteristics of this market we can grasp and use to further the understanding of financial markets in general. For this purpose we calculate minimal spanning trees for Warsaw's stock market for every studied year \cite{Mantegna:1997}.

If we are analysing financial markets as complex networks then it's natural to ask which stocks are most important in the network. The problem of defining and computing the importance of nodes in a graph is well defined in literature \cite{Newman:2003} and we use Markov centrality \cite{White:2003} to calculate the importance of every stock on the Warsaw's market in the minimally spanning trees, therefore we are able to see which stocks and particularly which sectors are influential on the Warsaw's stock market over time.

Researchers in social sciences (much like their counterparts in natural sciences) are often interested in crucial questions about predictability, which (also in economics and finance) are not trivial due to the human involvement \cite{Rosser:2008}. While researchers in the social sciences are interested in predictability of many processes, such as movements of people and their communication patterns \cite{Song:2010} the economists and financial researchers are most interested in the predictability of prices and their changes, due to the importance of prices in economics and finance and due to the vast amounts of data which can be analysed in this effort. We propose that question of predictability is more fundamental and is crucial to the problem of profitability of trading algorithms, and is indeed providing a way to analyse how close the markets are to the efficient market hypothesis, as under this hypothesis the price changes should be completely random and therefore unpredictable. We are therefore estimating predictability of price changes by estimating entropy rate of the relevant time series. This allows us to see how close the specific stocks and sectors are to the efficient market hypothesis at different times.

We are thus looking at the structure of the Warsaw's stock market and its changes to see how it has evolved over the years and what are the interesting characteristics of the market. We are particularly interested in how the structure of the Warsaw's stock market has been affected by the crisis of 2007-8. The structure of the Warsaw's market has been studied using the network approach \cite{Sienkiewicz:2013}, the scope of that study was different however, focusing on the global structure of the tree itself. Other studiee of the behaviour of Warsaw's stock market have been performed around the time of the 2007-8 crisis, do not use sufficiently developed methodology or large enough material to arrive at interesting conclusions \cite{Li:2008,Echaust:2013,Karpio:2013}. Some studies suggest that Warsaw's stock market is not fully efficient \cite{Kompa:2009}, perhaps the network approach can shed a little more light on the degree to which stocks in Warsaw adhere to the efficient market hypothesis. Thefore we see a need to study the structural changes of the Warsaw's stock market, particularly around the times of the recent financial crisis.

\section{Material and methods}

\begin{figure}[tbh]
\centering
\includegraphics[width=0.5\textwidth]{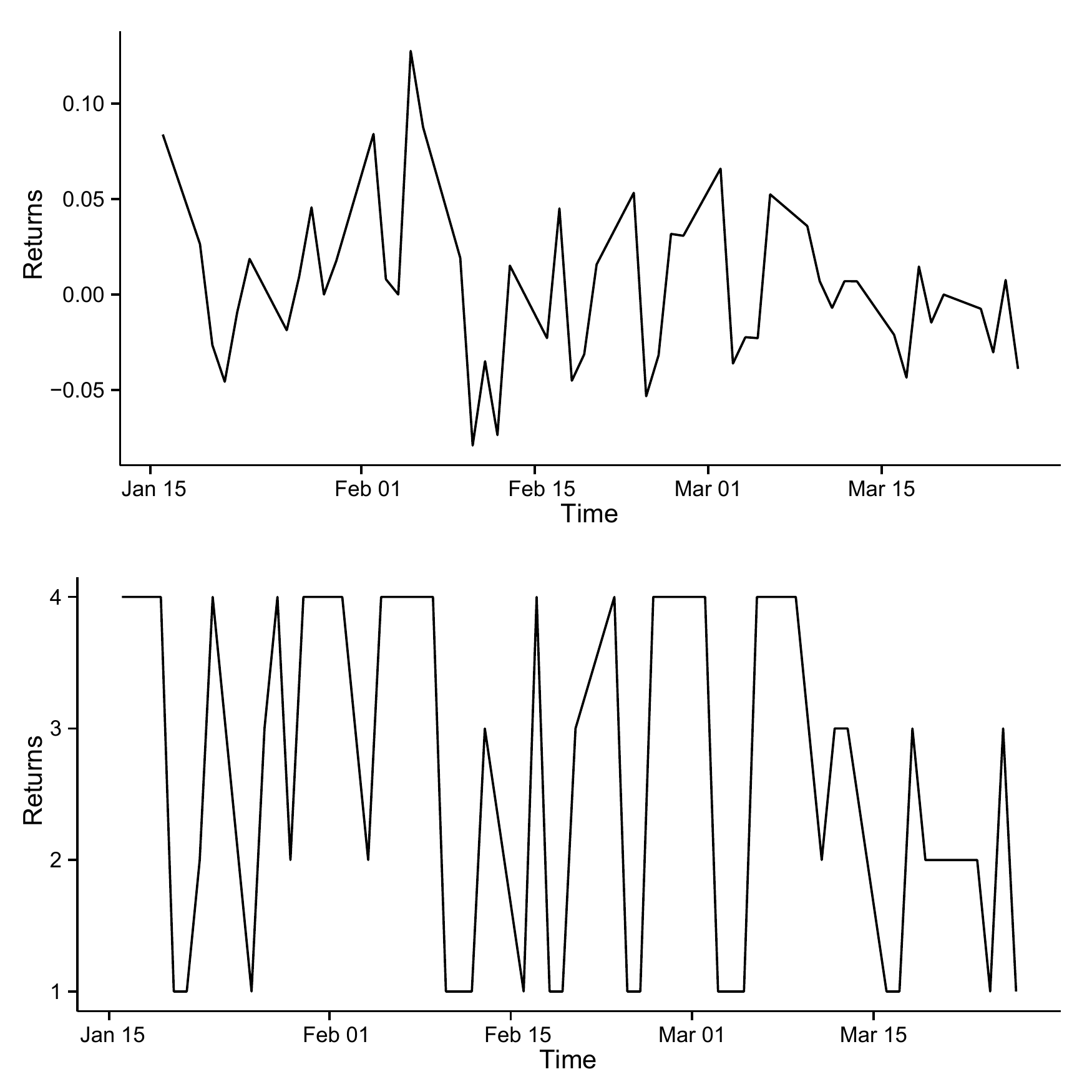}
\caption{Example of standard and discrete log returns}
\label{fig:digs}
\end{figure}

In this study we use daily price data for GPW (Warsaw's stock market -- Giełda Papierów Wartościowych) available at \url{http://bossa.pl/notowania/metastock/} for stocks listed on Warsaw's market between the 3rd of Janurary 2000 and the 5th of July 2013. Out of this dataset we choose 357 stocks which were listed during at least 1000 consecutive days. Those stocks are then divided into 26 sectors according to the Warsaw's market classification. We then divide those time series into yearly subseries, and for each year disregard the stocks with incomplete time series. Therefore we end up with the numbers of studied companies for every year which are shown in Table \ref{tab:numbers}. The data is transformed in the standard way for analysing price movements, that is so that the data points are the log ratios between consecutive daily closing prices:
\begin{equation}
r_{t}=ln(p_{t}/p_{t-1}).
\end{equation}
We use time series consisting of those log returns to calculate dependency networks as well as average log returns and their standard deviation. For the purpose of estimating the entropy rate of those time series we discretise those data points into 4 distinct states. The states represent 4 quartiles, therefore each state is assigned the same number of data points (an example shown on Fig. \ref{fig:digs}). This design means that the model has no unnecessary parameters, which could affect the results and conclusions reached while using the data. This and similar experimental setups have been used in similar studies \cite{Navet:2008} (Navet \& Chen divided data into 8 equal parts instead of quartiles) and proved to be very efficient \cite{Steuer:2001}.

\begin{table}[htbp]
\caption{Cardinality of studied company sets}
\centering
\begin{tabular}{lr}
\textbf{Year} & \textbf{Cardinality} \\ \toprule
2000 & 113 \\
2001 & 101 \\
2002 & 76 \\
2003 & 93 \\
2004 & 106 \\
2005 & 134 \\
2006 & 177 \\
2007 & 206 \\
2008 & 223 \\
2009 & 264 \\
2010 & 277 \\
2011 & 259 \\
2012 & 194 \\
2013 & 245 \\ \bottomrule
\end{tabular}
\label{tab:numbers}
\end{table}

First we build stock correlation network for the presented log returns time series. This allows us to investigate the hierarchical structure present 
in a portfolio of $n$ stocks (see Table \ref{tab:numbers}) traded on Warsaw's stock market. The topological arrangement of the stocks is based on the Pearson's correlation coefficient of difference of logarithms of closing prices for two consecutive days. This correlation coefficient is computed for all the pairs of stocks in the studied set.

The Pearson's correlation coefficient mentioned above is defined as \cite{Feller:1971}:
\begin{equation}
\rho_{ij}=\frac{E(r_i r_j) - E(r_i)E(r_j)}{\sqrt{(E(Y_i^2)-E(Y_i)^2)(E(Y_j^2)-E(Y_j)^2)}}
\end{equation}
where $i$ and $j$ are the labels of stocks, and $r_i$ is similar to $r_t$ specified above, only calculated for stock $i$ for a given day $t$. The average is then computed for all trading days in the investigated period (over all $t$).

From the properties of correlation we know that this matrix is symmetric, with the diagonal being filled with $\rho_{ii}=1$. Thefore each set we investigate consists of $n~(n-1)/2$ meaningful correlation coefficients \cite{Mantegna:1997}. The correlation coefficient of a pair of stocks is not an Euclidean metric, so we can't use it for the network topology. Therefore a generalised metric is defined based on the above-mentioned correlation, to find an approximate distance between the two stocks. Usually the below is used:
\begin{equation}
d(i,j)=1-\rho_{ij}^2.
\end{equation}

This function guarantees that $d(i,j)$ is an Euclidean metric, that is:
\begin{enumerate}
\item $d(i,j)=0$ if and only if $i=j$;
\item $d(i,j)=d(j,i)$;
\item $d(i,j) \le d(i,k) +d(k,j)$.
\end{enumerate}

The distance matrix $D$ is then used to determine the minimal spanning tree \cite{Papa:1982} connecting the $n$ stocks in the portfolio. From the distance matrix $D$ we create a list $S$ by ordering the distances in decreasing order. Then starting from the first element of the list we add the corresponding link if and only if the resulting graph is still a forest or a tree \cite{Aste:2005}. Similarly a planar maximally filtered graph can be constructed by adding the corresponding link if and only if the resulting graph is still planar (genus equal $0$).

\begin{figure}[tbh]
\centering
\includegraphics[width=\textwidth]{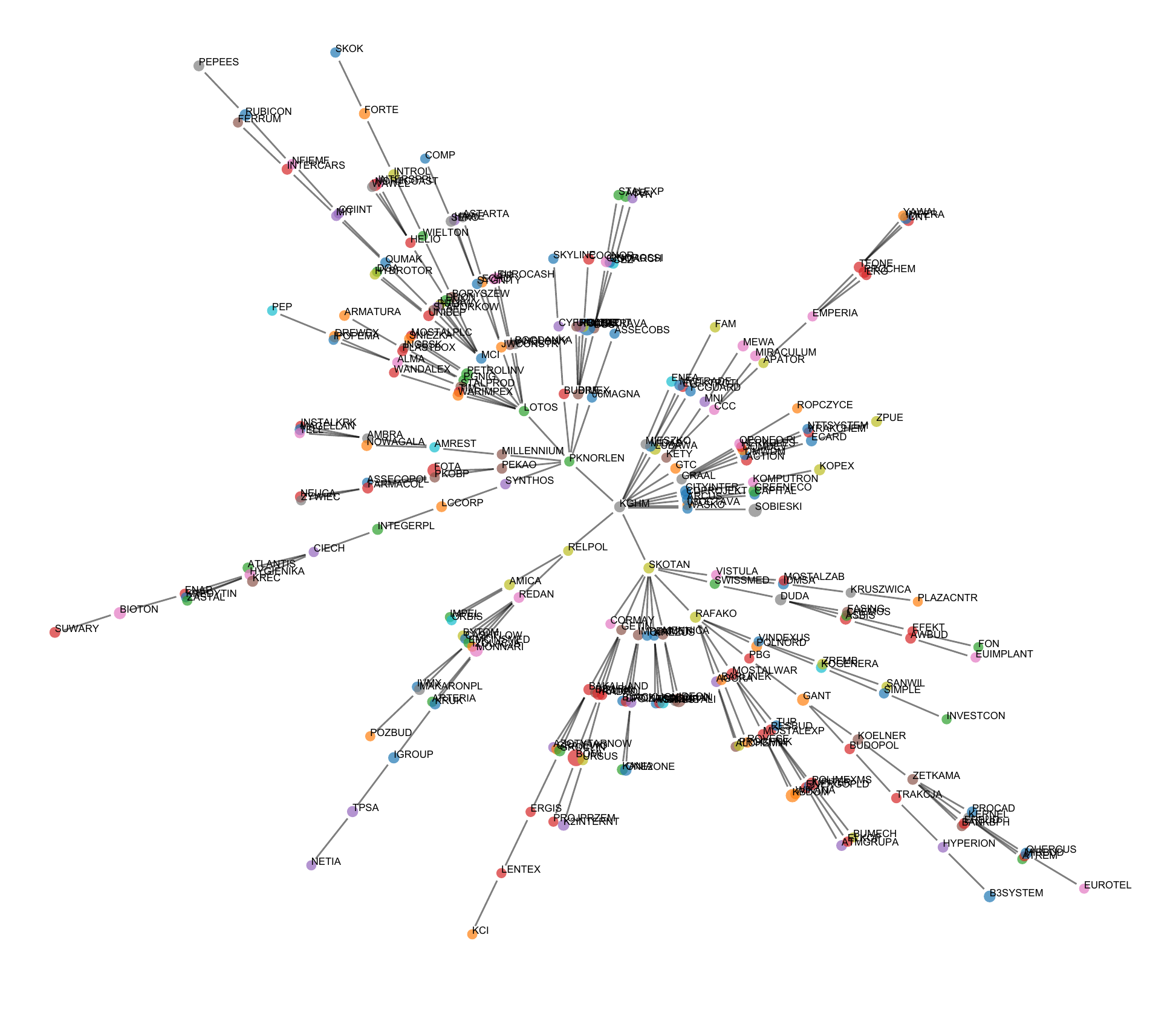}
\caption{{Minimal spanning tree for log returns (WIG, 2013)}}
\label{fig:mst2013}
\end{figure}

The minimal spanning tree provides a very topologically restrictive arrangement of stocks, which selects the most relevant connections of each point of the set. Therefore the hierarchical organisation found this way is highly interesting from an economic point of view, providing the most important dependencies on the market, which has been shown in numerous studies. An example of the tree created for 2013 is presented on Fig. \ref{fig:mst2013} (size of the node corresponds to the average log returns, and width of the edges is determined by correlation between the connected nodes). We are not using other structures proposed in literature besides minimal spanning tree, such as planar maximally filtered graph \cite{Lillo:2010}, because of our large portfolios we want a structure which is highly restrictive and only presents the most relevant connections, therefore we use the highly topologically restrictive minimal spanning tree.

One of the most important metrics in any network is centrality. Central nodes in a graph are often seen as the important agents, through which the interactions are conducted (be it social interactions, economic processes or biological interactions). Centrality is a good indicator of the relative popularity of individual nodes \cite{Faust:1994}. There are numerous ways to quantify centrality of a network \cite{Newman:2003}, their quality is usually comparable, hence we have chosen Markov centrality due to its general nature.

For any undirected graph $G=\{V,E\}$ where $V$ denotes a set of nodes and $E$ denotes a set of edges centrality is a function $C: V \rightarrow \mathcal{R}$. Centrality cannot be a function of discrete set of nodes, therefore it's a function of simpler quality such as the degree of a node $v_i$ denoted as $\Psi(v_i)$. Then centrality forms a matrix:
\begin{equation}
c_{ij}=\frac{\partial C(v_i)}{\partial \Psi_{j}}
\end{equation}
which indicates the influence of a change in degree of node $v_j$ on the centrality of the node $v_i$, which can be positive (node $v_j$ improves the importance of the node $v_i$ if an edge would be added to it) or negative (the opposite) \cite{Jeong:2001}.

Markov centrality interprets the network as a Markov process and can be intuitively understood as the amount of time a token performing a random walk spends on each node. This can be computed as the mean first-passage time in the Markov chain:
\begin{equation}
m_{rt}=\sum_{n=1}^{\infty}{nf_{rt}^{(n)}}
\end{equation}
where $f_{rt}^{(n)}$ denotes the probability that the chain first returns to the node $t$ in exactly $n$ steps. This can be computed as a matrix:
\begin{equation}
M=(I-Z+EZ_{dg})D
\end{equation}
where $I$ is the identity matrix, $E$ is a matrix filled with ones, and $D$ is a diagonal matrix is filled with reciprocal of the stationary distribution $\pi(v)$ of a node $v$. $Z_{dg}$ is the diagonal of the fundamental matrix of the Markov chain \cite{White:2003}.

Markov centrality is then defined as \cite{Brandes:2005}:
\begin{equation}
c_{M}(v)=\frac{n}{\sum_{s \in V} M_{sv}}
\end{equation}
where $n$ is the number of nodes.

We also calculate predictability of the discrete time series described above. The predictability of a time series can be estimated using the entropy rate. For a stationary stochastic process $X = \{X_i\}$, the entropy rate is defined as
\begin{equation}
	\label{eq:Def_entropy_rate}
	H(X) = \lim_{n \rightarrow \infty} \frac{1}{n} H(X_1, X_2, \dots, X_n)
\end{equation}
The right side of \eqref{eq:Def_entropy_rate} can be interpreted such as that entropy rate measures the uncertainty in a quantity at time $n$ having observed the complete history up to that point. Theory of information defines entropy rate of a stochastic process as the amount of new information created in a unit of time \cite{Cover:1991}. Hence entropy rate can be interpreted as maximum rate of information creation which can be processed as price changes.

Methods of entropy estimation can be grouped into two separate categories \cite{Gao:2006}: maximum likelihood estimators and estimators based on data compression algorithms. The latter are considered better suited for economic data and are precise even for short time series \cite{Louchard:1997,Leonardi:2010}.

We use an estimator based on Lempel-Ziv algorithm, which measures complexity in Kolmogorov's sense \cite{Cover:1991}. On the basis of Lempel-Ziv algorithm \cite{Lempel:1977} there have been a number of estimators of entropy rate created. In this article we follow \cite{Navet:2008} and use the estimator created by Kontoyiannis in 1998 (estimator $a$) \cite{Kontoyiannis:1998}. This estimator is widely used \cite{Kennel:2005,Navet:2008} and it was shown that it has better statistical properties than previous estimators based on Lempel-Ziv algorithm \cite{Kontoyiannis:1998}, though there is a large choice of slightly different variants to choose from \cite{Gao:2008}, which is largely irrelevant.

More formally, to calculate the entropy of a random variable $X$, the probability of each possible outcome $p(x_i)$ must be known. When these probabilities are not known, entropy can be estimated by replacing the probabilities with relative frequencies from observed data. Estimating the entropy rate of a stochastic process is more complex as random variables in stochastic processes are usually interdependent. Then the mentioned estimator is defined as:
\begin{equation}
	\label{eq:LZ_complexity}
	\hat{H_{lz}} = \frac{n \log_2 n}{\sum_i \Lambda_i},
\end{equation}
where $n$ denotes the length of the time series, and $\Lambda_i$ denotes the 
length of the shortest substring starting from time $i$ 
that has not yet been observed prior to time $i$, i.e. from time $1$ to 
$i-1$. 
It is known that for stationary ergodic processes, $\hat{H_{lz}}(X)$ converges to the entropy rate $H(X)$ almost surely as $n$ approaches infinity \cite{Kontoyiannis:1998}.

Figures \ref{fig:avcentr}, \ref{fig:agcentr}, \ref{fig:avlogr}, \ref{fig:avstdev} \& \ref{fig:avlz} use two-tone pseudo colouring (also called Horizon graphs) as a tool for  compact visualization for many sectors. Basically a line chart is layered into 2 layers for positive values and 2 layers for negative values. The second layer is stacked upon the first (second layer appears darker), and the negative values are superimposed on the same band (handing from the top of it), providing a space-efficient time series visualisation technique \cite{Miyamura:2005,Heer:2009}.

Figures \ref{fig:marketret}, \ref{fig:marketweiret}, \ref{fig:marketstdev}, \ref{fig:marketlz}, \ref{fig:marketweisd} \& \ref{fig:marketweilz} use local polymonial regression fitting for the trend lines, also showing 95\% confidence bands \cite{Chambers:1991}.

Figures \ref{fig:lzstdev} \& \ref{fig:lzstdev100} use linear regression (least squares estimator) for the visible regression lines.

\section{Results}

All the results have been calculated for specific companies for specific years between 2000 and 2013. The results are then grouped by sector and grouped for the whole market as well, which divides the discussion into the two corresponding grous.

First we take a look at the results aggregated into sectors. Warsaw's stock market groups the listed companies into 26 specific sectors and this split has been used.

\begin{figure}[tbh]
\centering
\includegraphics[width=0.5\textwidth]{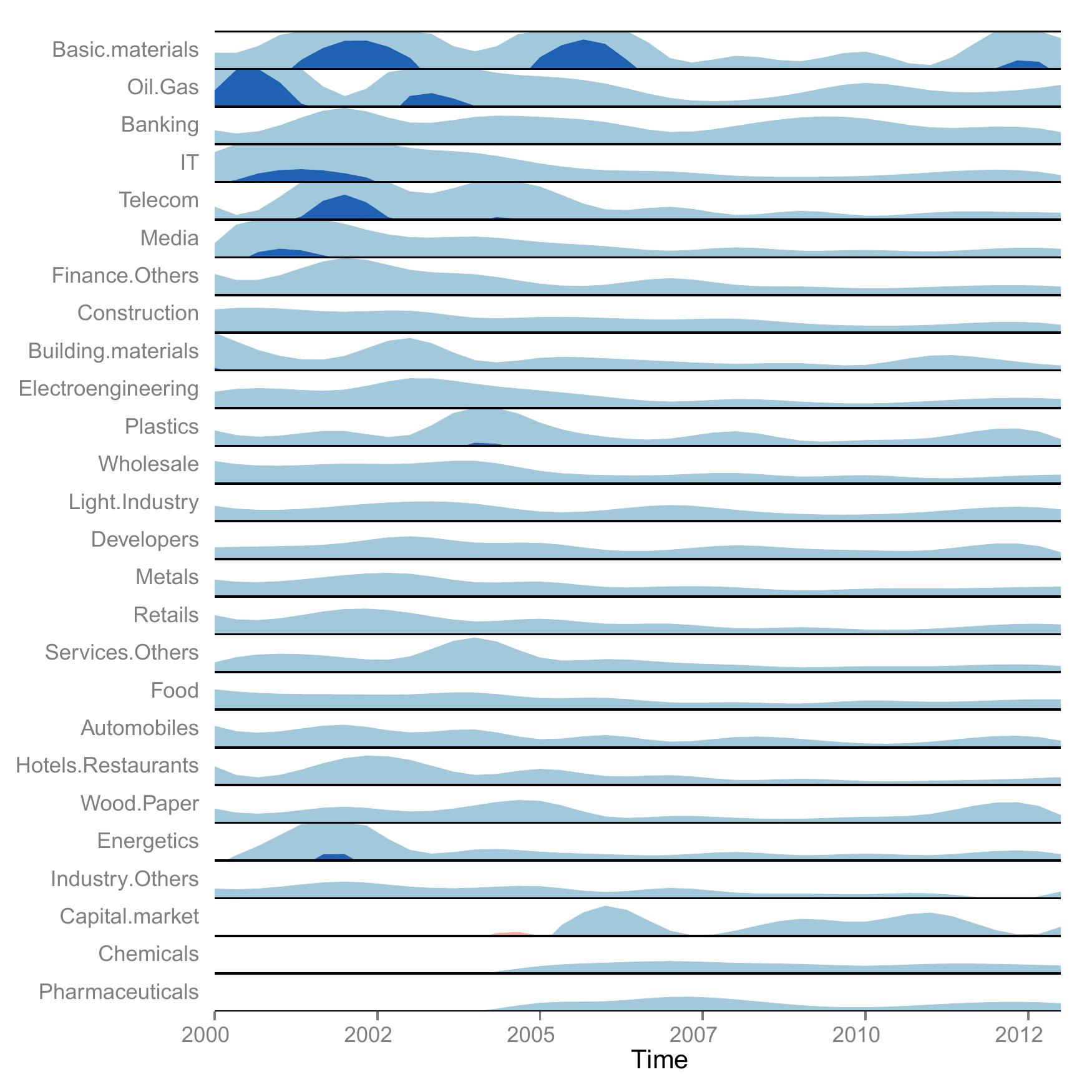}
\caption{{Average Markov centrality by sector}}
\label{fig:avcentr}
\end{figure}

\begin{figure}[tbh]
\centering
\includegraphics[width=0.5\textwidth]{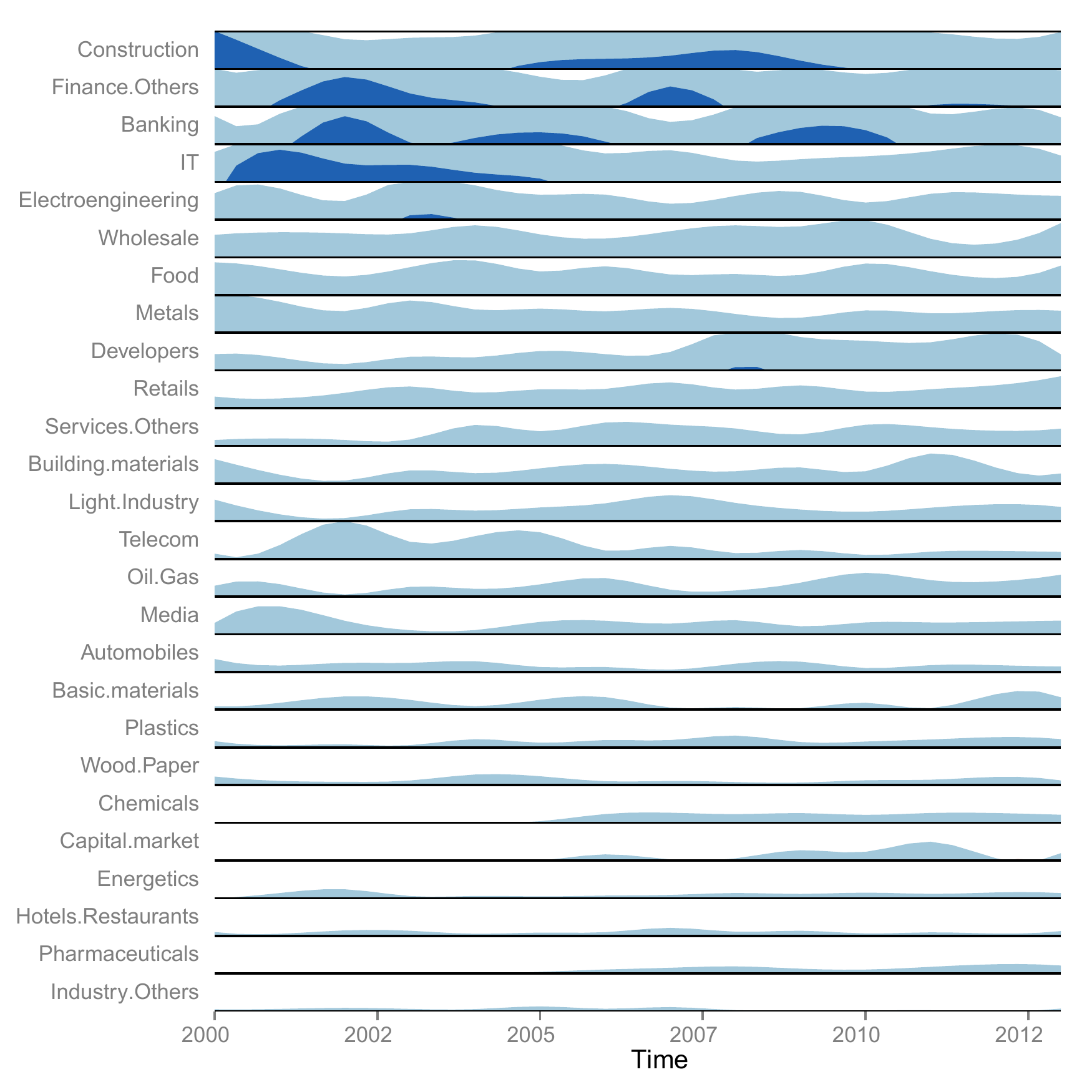}
\caption{{Aggregate Markov centrality by sector}}
\label{fig:agcentr}
\end{figure}

On Fig. \ref{fig:avcentr} the average Markov centrality by sector for the studied years is presented, while on Fig. \ref{fig:agcentr} we have presented the aggregate Markov centrality by sector for the same period. Both measures are obviously related but while the average value shows the changes happening in the whole sector, the aggregate value shows more clearly the impact of a given sector to the whole market, but loses sight of the changes happening to the whole sector and may be dependent on a couple important stocks instead. This distinction is thus similar to the distinction between mean and median.

The crisis of 2007-08 has been called financial, but it hasn't been the whole of the financial industry which lead to it and instead it has mostly been the banking sector \cite{Turkes:2010}. We will therefore focus our attention on the banking sector in this analysis to see how has the banking sector been affected and affecting the whole market in the last few years.

On Fig. \ref{fig:avcentr} we can see that in the time of crisis the average Markov centrality of most sectors has dwindled, that is to say the market has become more diffused. But at the same time the importance of banking sector and also to a degree the capital markets sector have risen. The same situation can be seen on Fig. \ref{fig:agcentr}. Banking becomes the central sector of the Warsaw's market in the period of 2008-2010. This shows that the negative processes within the banking industry which led to the financial crisis have been spreading into the whole market, therefore degrading the well-being of the whole of Warsaw's market.

But we can also see that in the period of 2012-2013 the banking's sector influece has steadily fallen, and the market is yet again driven by industries related to the real economy, that is basic materials and oil \& gas in terms of average Markov centrality, and construction together with wholesale, food, and other financial institutions in terms of aggregate Markov centrality. This analysis allows us to conclude that the financial crisis, if we understand it as the negative influence of the banking sector on the processes taking place on Warsaw's stock market, has come to an end in 2013. This does not mean that the economic situation is perfect, but it does suggest that the market has healed itself from the wounds inflicted by the banking industry back in 2007 and can now return to its normal functioning, potentially leading to better economic prospects.

\begin{figure}[tbh]
\centering
\includegraphics[width=0.5\textwidth]{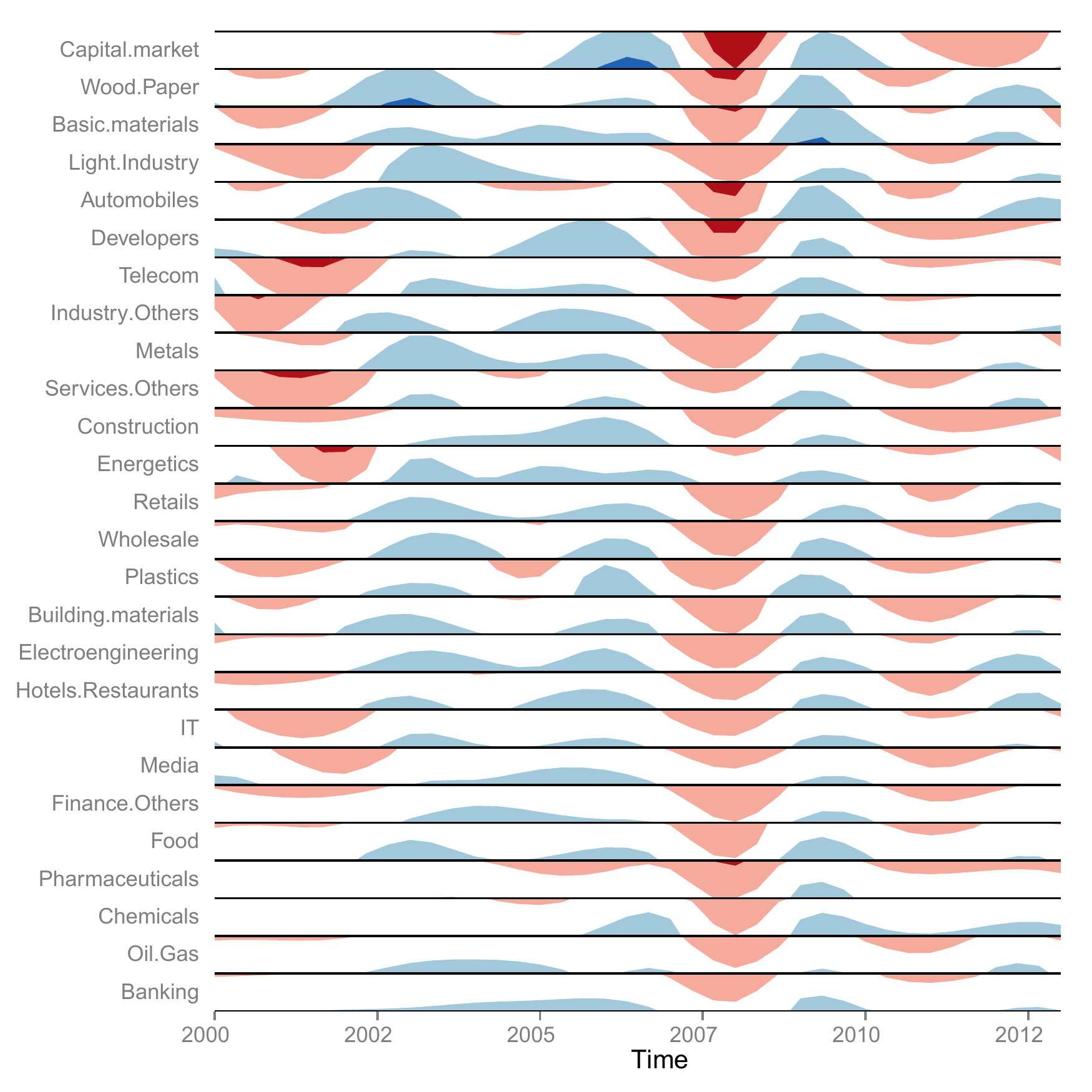}
\caption{{Average log returns by sector}}
\label{fig:avlogr}
\end{figure}

\begin{figure}[tbh]
\centering
\includegraphics[width=0.5\textwidth]{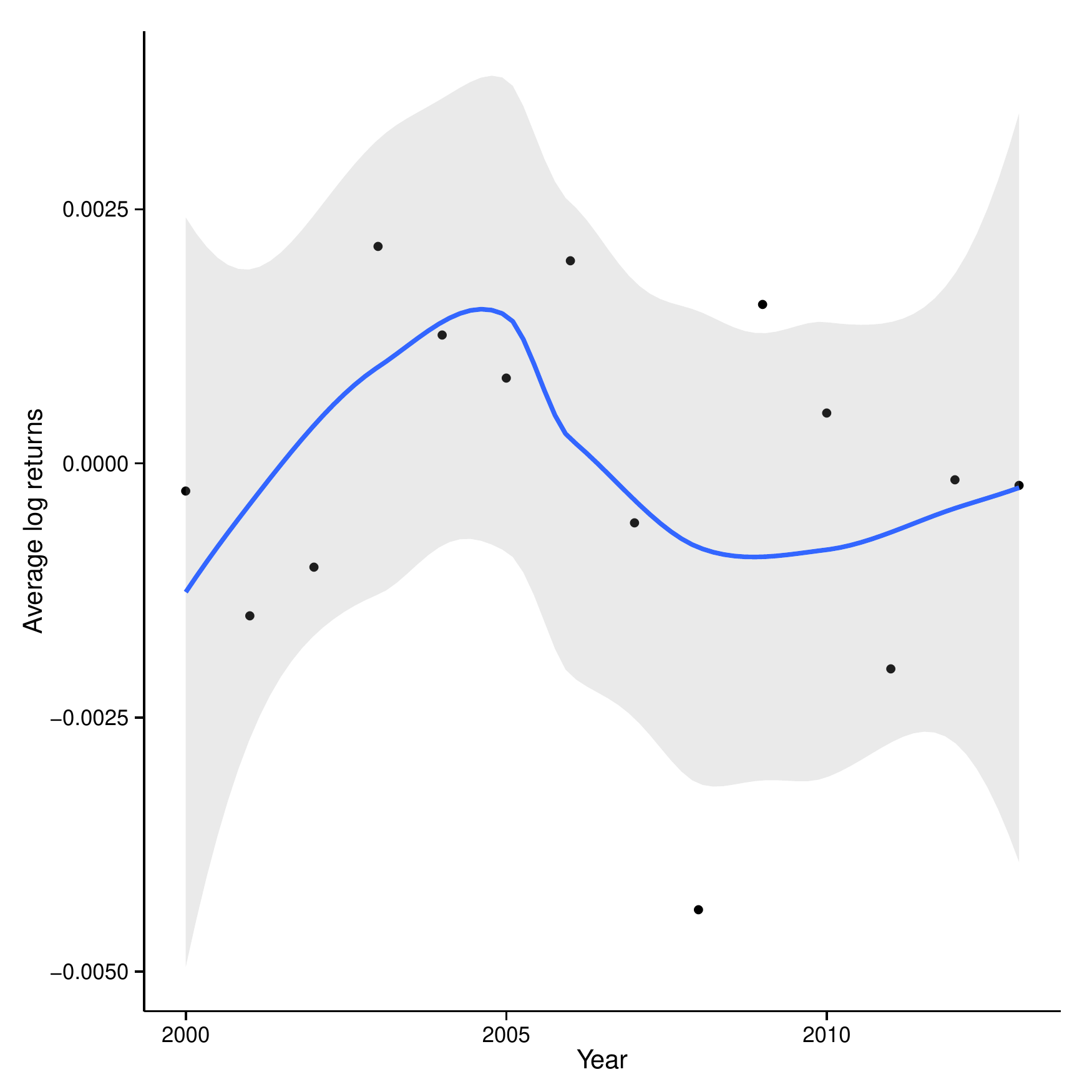}
\caption{{Average log returns}}
\label{fig:marketret}
\end{figure}

The last point is well illustrated by Fig. \ref{fig:avlogr}. We do not by at all mean that the market is growing steadily in 2013. In fact we can see that while in the middle of the crisis in 2009 the markets have performed better than in the later years. While 2012 has been a relatively good year the 2013 appears to be worse than 2012 in terms of average log returns in most sectors. The outlook is not as bad as it seems from this picture, which can be observed on Fig. \ref{fig:marketret} to which we'll return in our analysis. Instead we meant that the structural changes in the market itself promote the idea that the market is breaking free from the situation it had been caught by in 2007-8, which should lead to healthier operations on Warsaw's stock market in the near future.

\begin{figure}[tbh]
\centering
\includegraphics[width=0.5\textwidth]{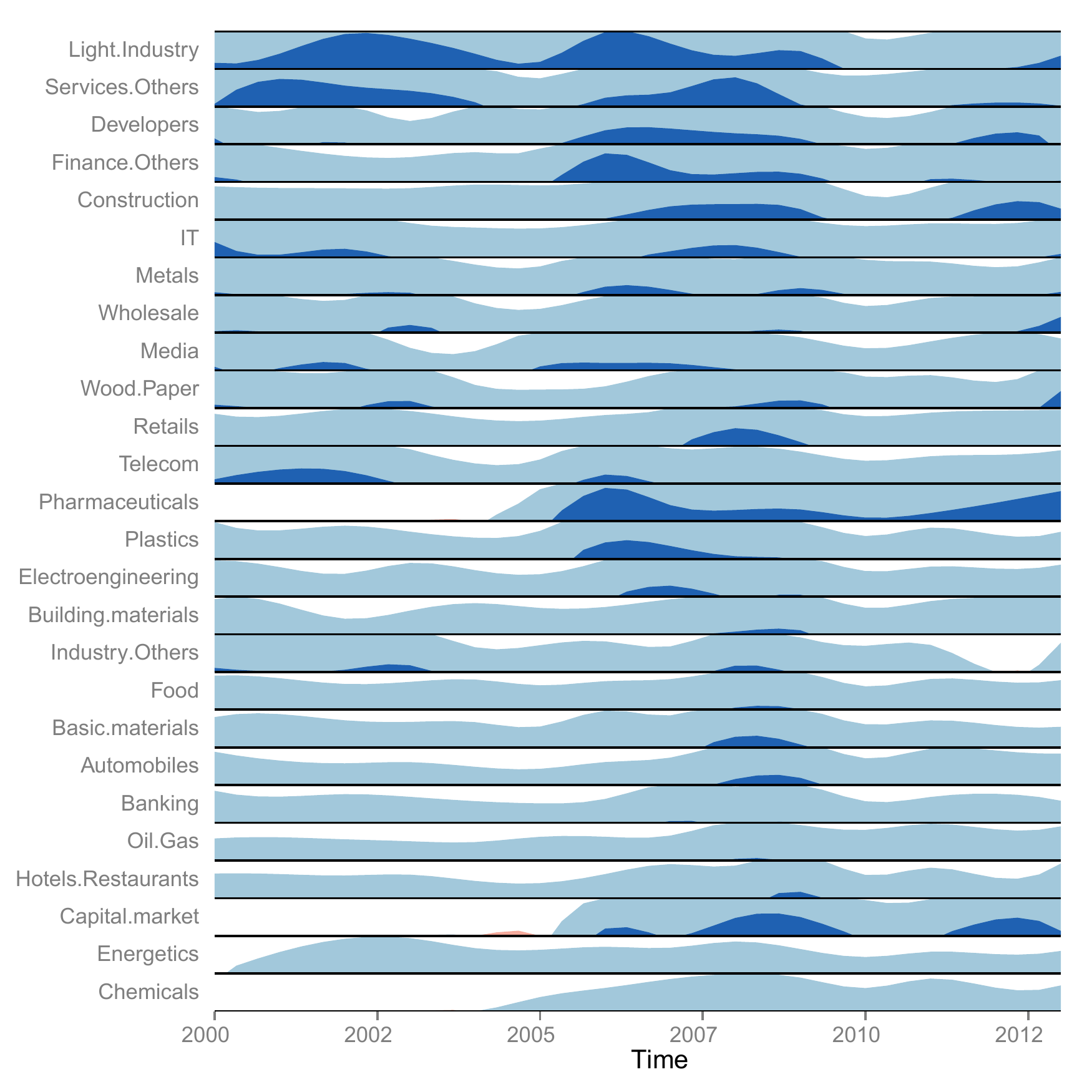}
\caption{{Average SD of log returns by sector}}
\label{fig:avstdev}
\end{figure}

We have analyzed the first moment of the data, it may prove insightful to analyse the second central moment as well. We present the average standard deviation of log returns divided by sector on Fig. \ref{fig:avstdev}. This way we can see how dispersed the log returns are in different sectors and how it changed over the years. It appears that the financial crisis has caused the average standard deviation in most industries to fall in the period of 2008-2010 (with services and capital markets seeing the biggest decline). This may hint that the market has been functioning in a manner which does not allow much diversity, therefore a manner which does not allow the market to be efficient either. If there is no diversity in opinion about the price then the market forces are not working efficiently at deriving the price. We can see that since 2010 the standard deviation of log returns has been increasing (with the biggest in pharmaceuticals, wholesale and light industry), showing perhaps that the market is getting back to healthy operations with diverse market forces.

\begin{figure}[tbh]
\centering
\includegraphics[width=0.5\textwidth]{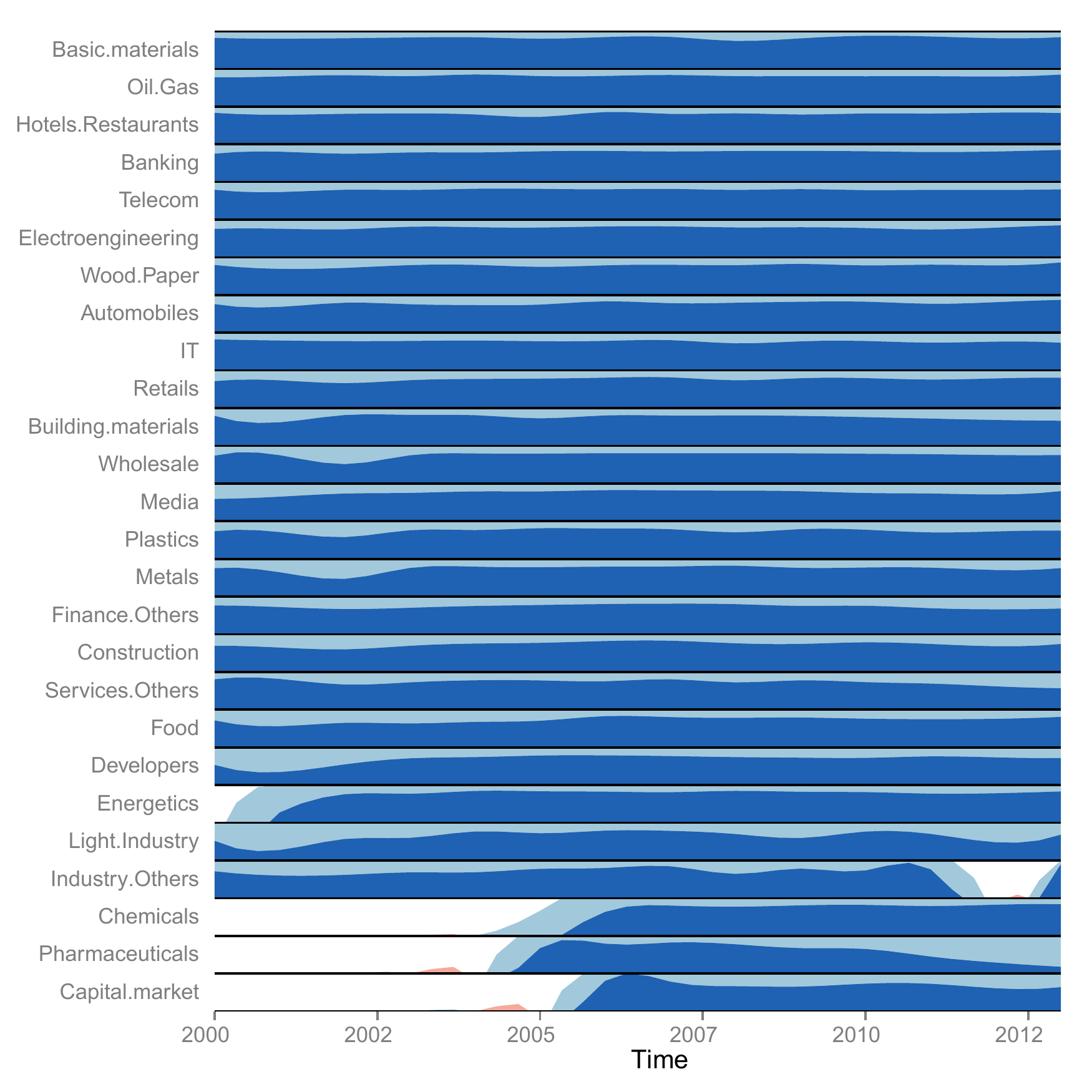}
\caption{{Average entropy rate by sector}}
\label{fig:avlz}
\end{figure}

Besides standard deviation we also look at average entropy rate calculated for the log returns which are presented on Fig. \ref{fig:avlz}. The entropy rate informs us about the predictability of the given price formation process, or in other words about the existence of patterns in the time series and the degree to which they are present in a given price formation process (the bigger the entropy rate of the process the closer it is to being an iid process and therefore closer to the efficient market hypothesis \cite{Fiedor:2013}). The predictability seems to be quite stable in time and not affected particularly by the financial crisis. Nonetheless while in 2007 most sectors had a very similar degree of predictability of the price changes the situation in 2013 is much more diverse(with pharmaceuticals and services becoming more predictable, while wood, paper and automobiles became less predicatble), showing that perhaps there has been a change to how the price formation processes assign significance to the sector to which a given stock belongs.

\begin{figure}[tbh]
\centering
\includegraphics[width=0.5\textwidth]{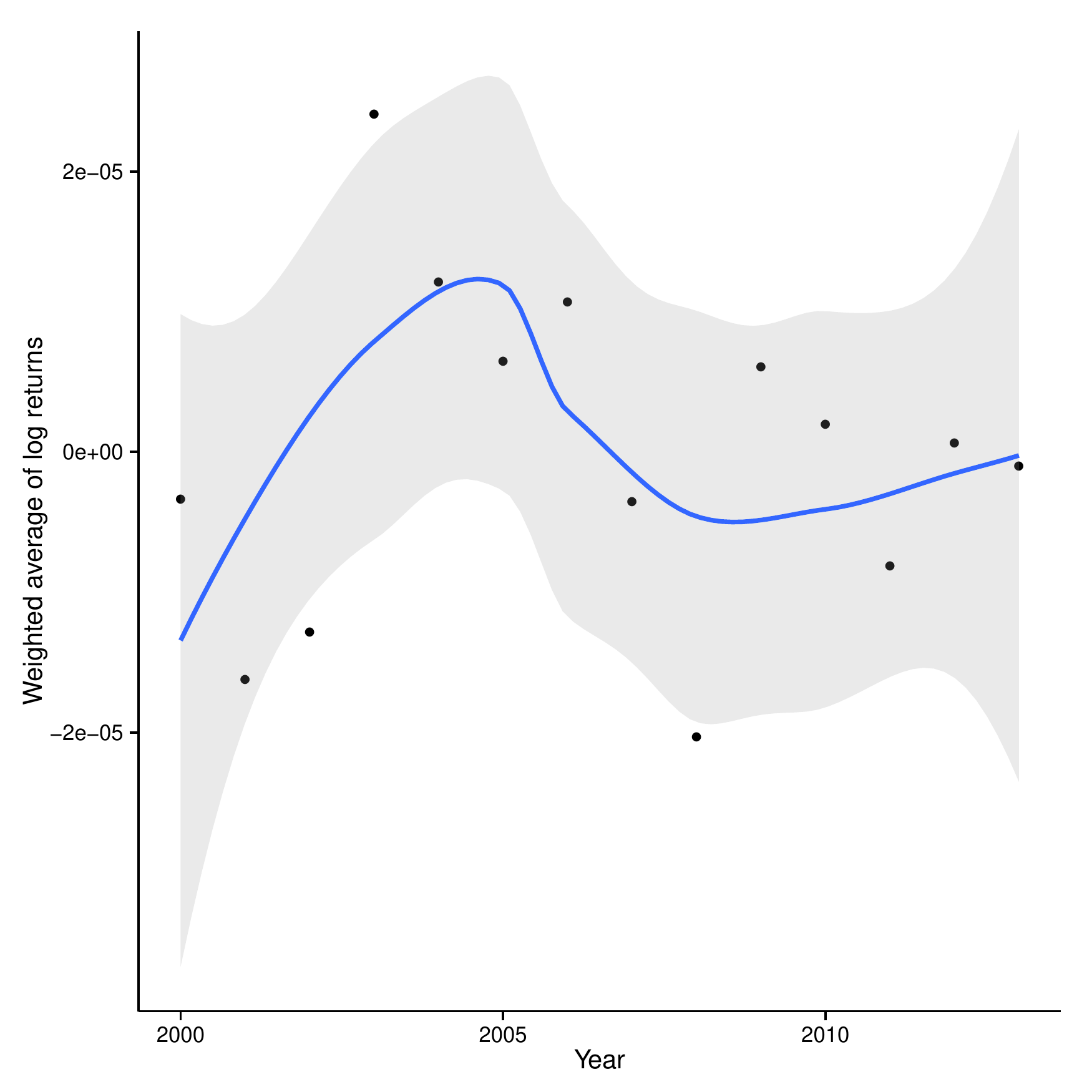}
\caption{{Weighted average log returns}}
\label{fig:marketweiret}
\end{figure}

Here we start a more aggregated analysis looking at the Warsaw's market as a whole. First we look at the average log returns for the whole market. Those values are shown on Fig. \ref{fig:marketret}. The same values weighted by the Markov centrality are presented on Fig. \ref{fig:marketweiret}. On those two we can see that despite our earlier point about the returns in 2013 not being very good the overall trend is positive since 2010, while it has been negative before. Therefore we can conclude that the outlook for Warsaw's stock market is relatively good.

\begin{figure}[tbh]
\centering
\includegraphics[width=0.5\textwidth]{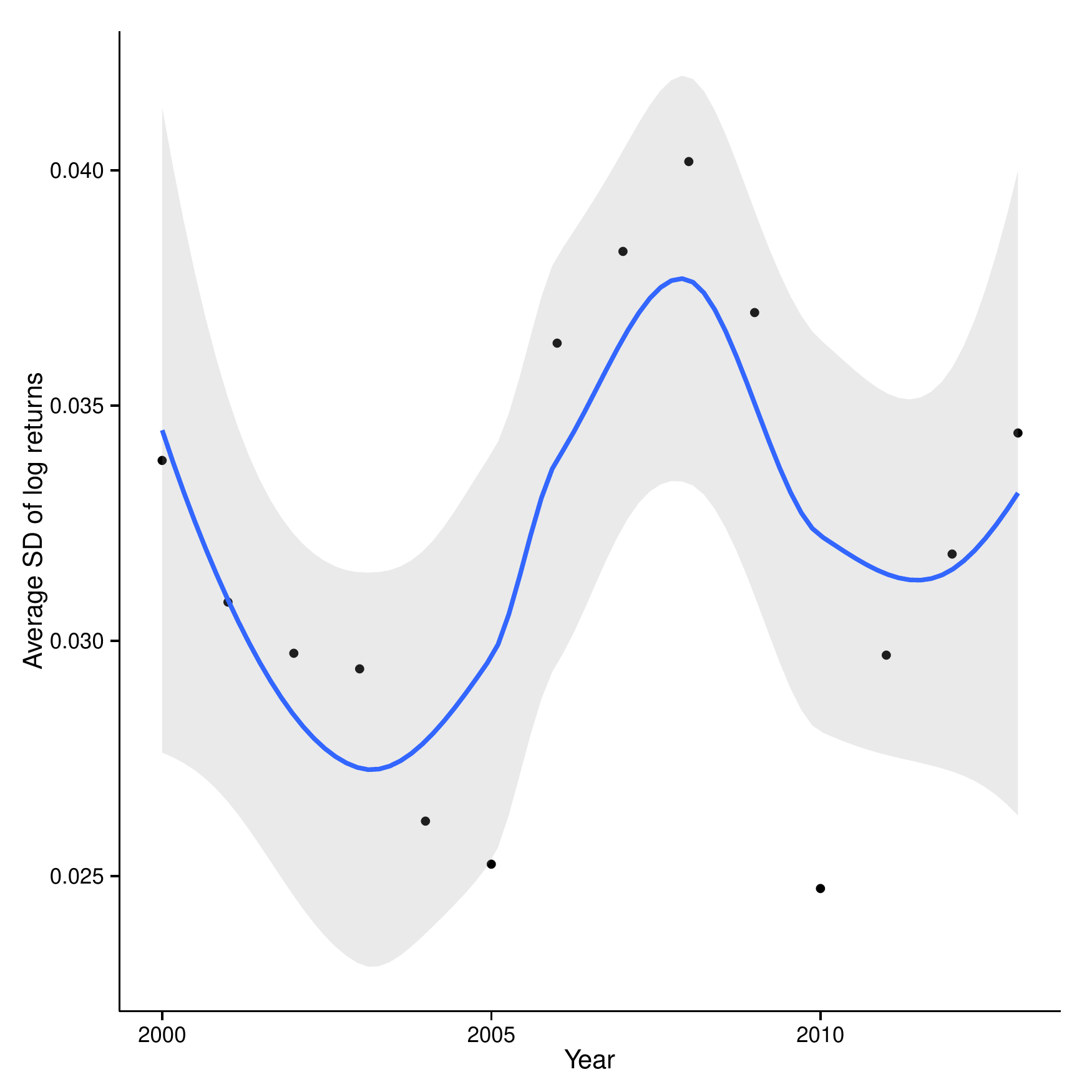}
\caption{{Average SD of log returns}}
\label{fig:marketstdev}
\end{figure}

\begin{figure}[tbh]
\centering
\includegraphics[width=0.5\textwidth]{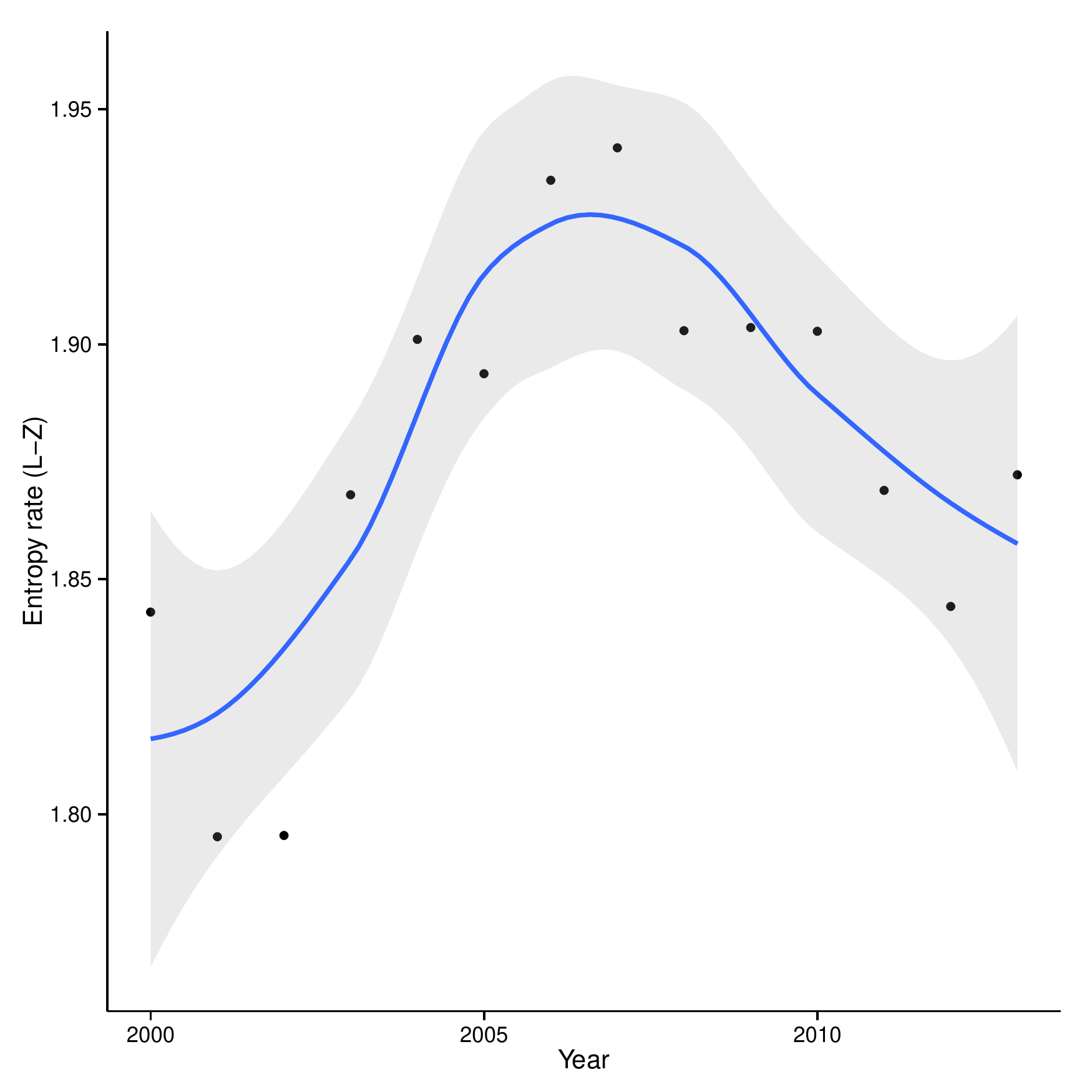}
\caption{{Average entropy rate}}
\label{fig:marketlz}
\end{figure}

On Fig. \ref{fig:marketstdev} we can see the confirmation that indeed the variability in prices changes dwindles during the crisis but is being regained since 2010, which is a good sign for the market in Warsaw. Interestingly if we look at the Fig. \ref{fig:marketlz} we can see that the average entropy rate for the whole market is behaving the opposite way to the standard deviation, that is it's steadily decreasing, and the price formation processes are becoming more predictable (the Pearson's correlation between the two is $-0.34$). This is surprising, as we would expect more variability in the markets to be bringing the market closer to the efficient market hypothesis, therefore the price formation processes closer to randomness.

\begin{figure}[tbh]
\centering
\includegraphics[width=0.5\textwidth]{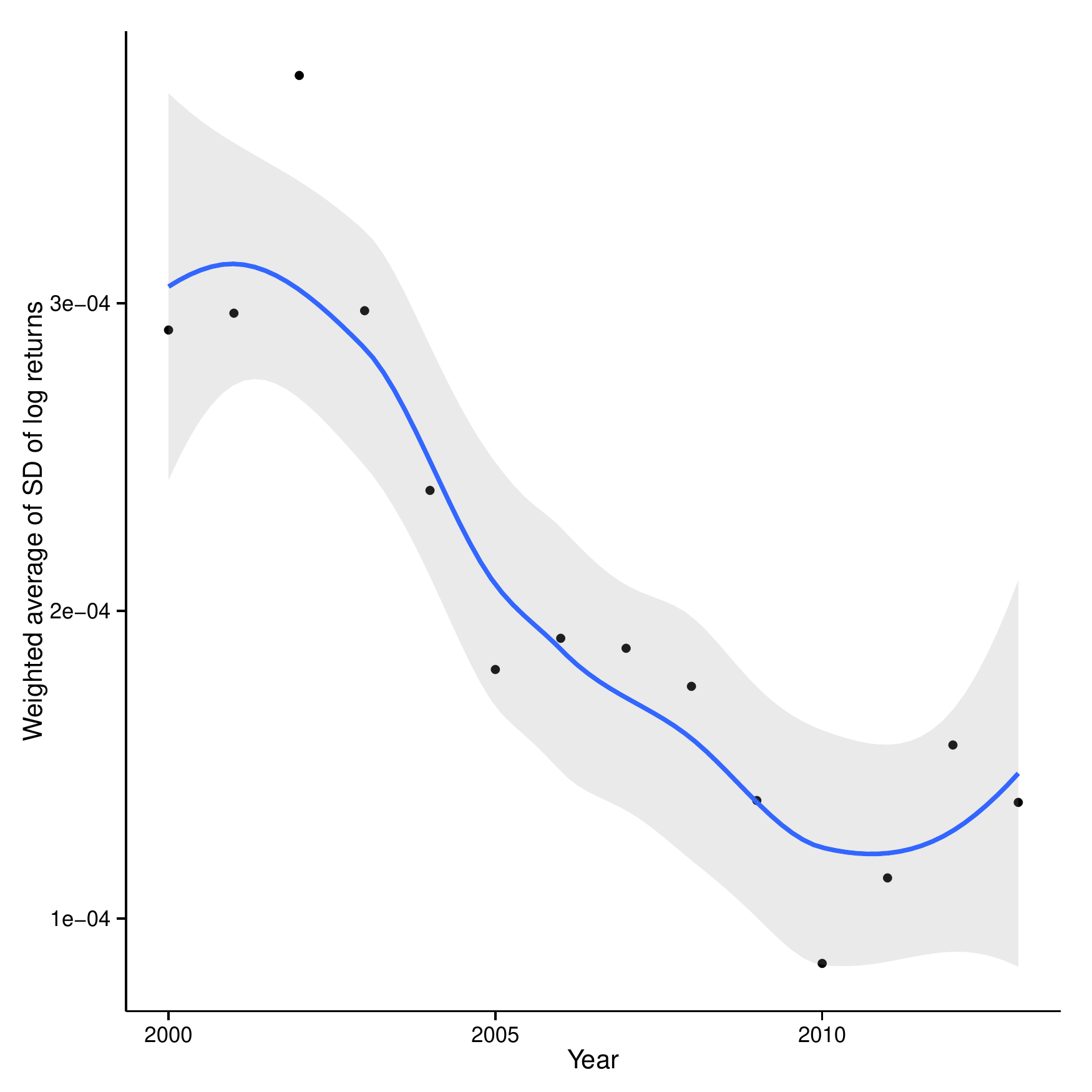}
\caption{{Weighted average SD of log returns}}
\label{fig:marketweisd}
\end{figure}

\begin{figure}[tbh]
\centering
\includegraphics[width=0.5\textwidth]{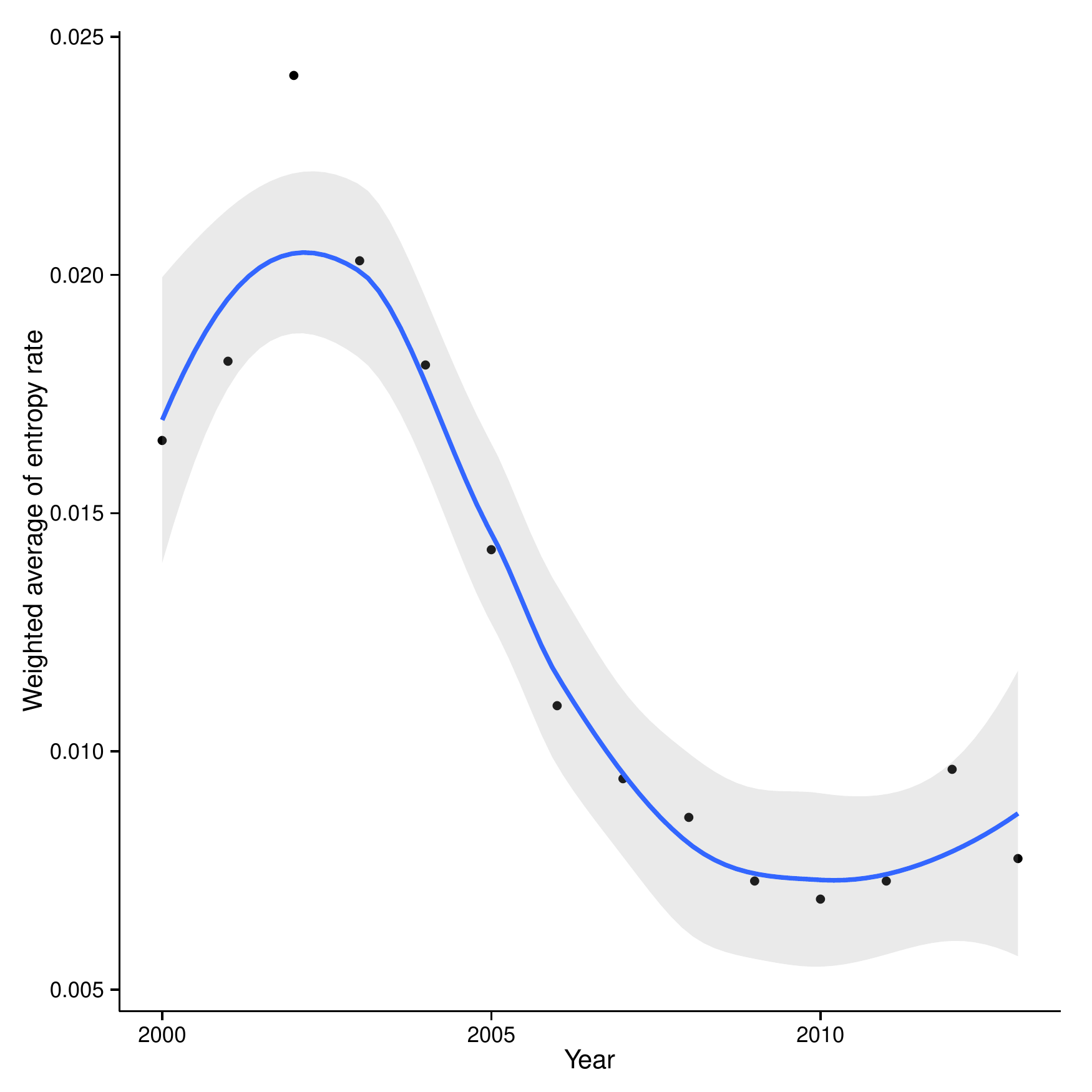}
\caption{{Weighted average entropy rate}}
\label{fig:marketweilz}
\end{figure}

On Figs. \ref{fig:marketweisd} and \ref{fig:marketweilz} we see the same values weighted by the Markov centrality. Here we see that both measures behave the way average unweighted log returns behave, therefore the surprising effect mentioned above has disappeared. It could therefore follow that the more important stocks are behaving the way we'd expect them to, while the more peripheral stocks are behaving counterintuitively. It may be that the less important stocks on the Warsaw's market do not have the trading volume to render them closer to the efficient market hypothesis.

\begin{figure}[tbh]
\centering
\includegraphics[width=0.5\textwidth]{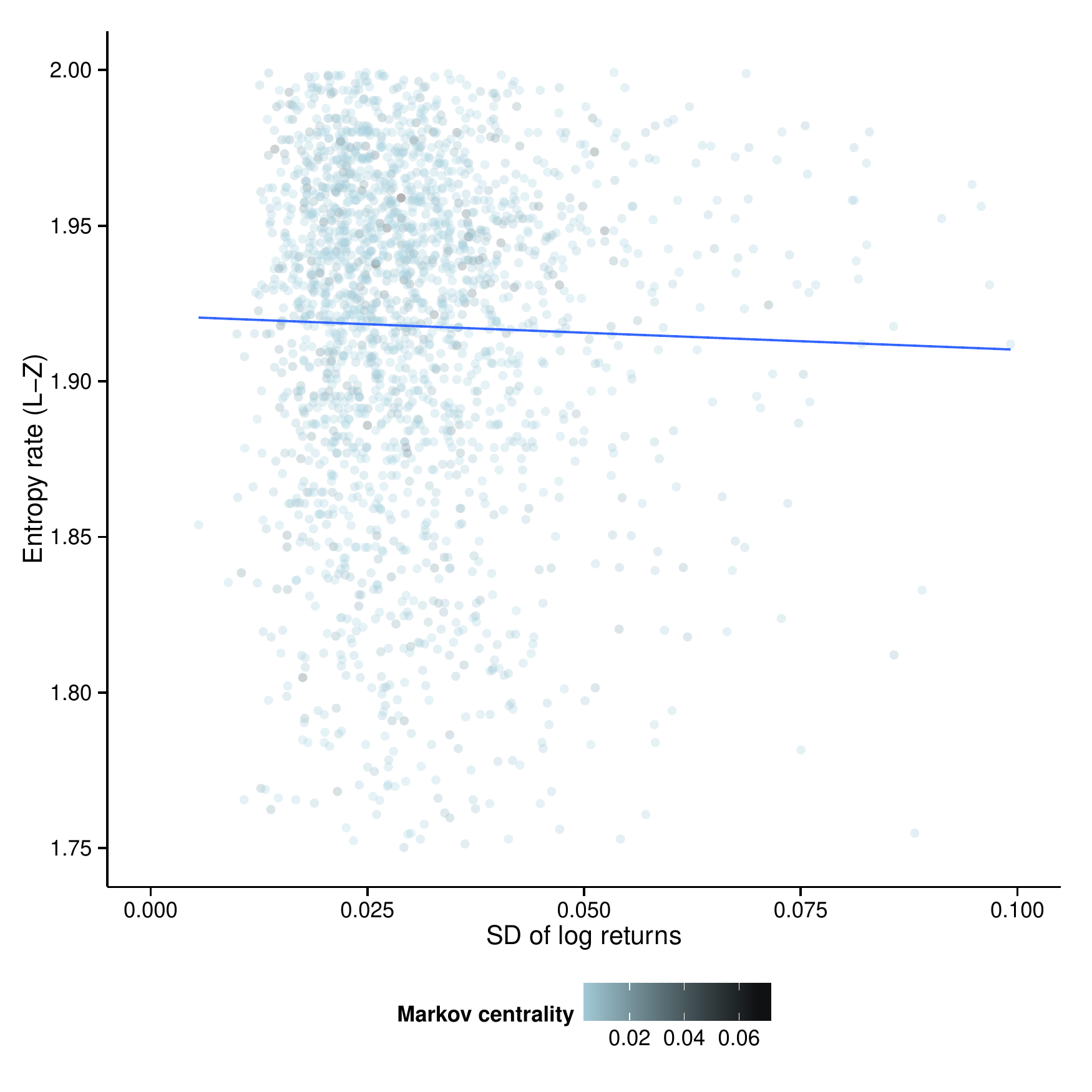}
\caption{{Entropy rate vs. SD of log returns (all datapoints)}}
\label{fig:lzstdev}
\end{figure}

\begin{figure}[tbh]
\centering
\includegraphics[width=0.5\textwidth]{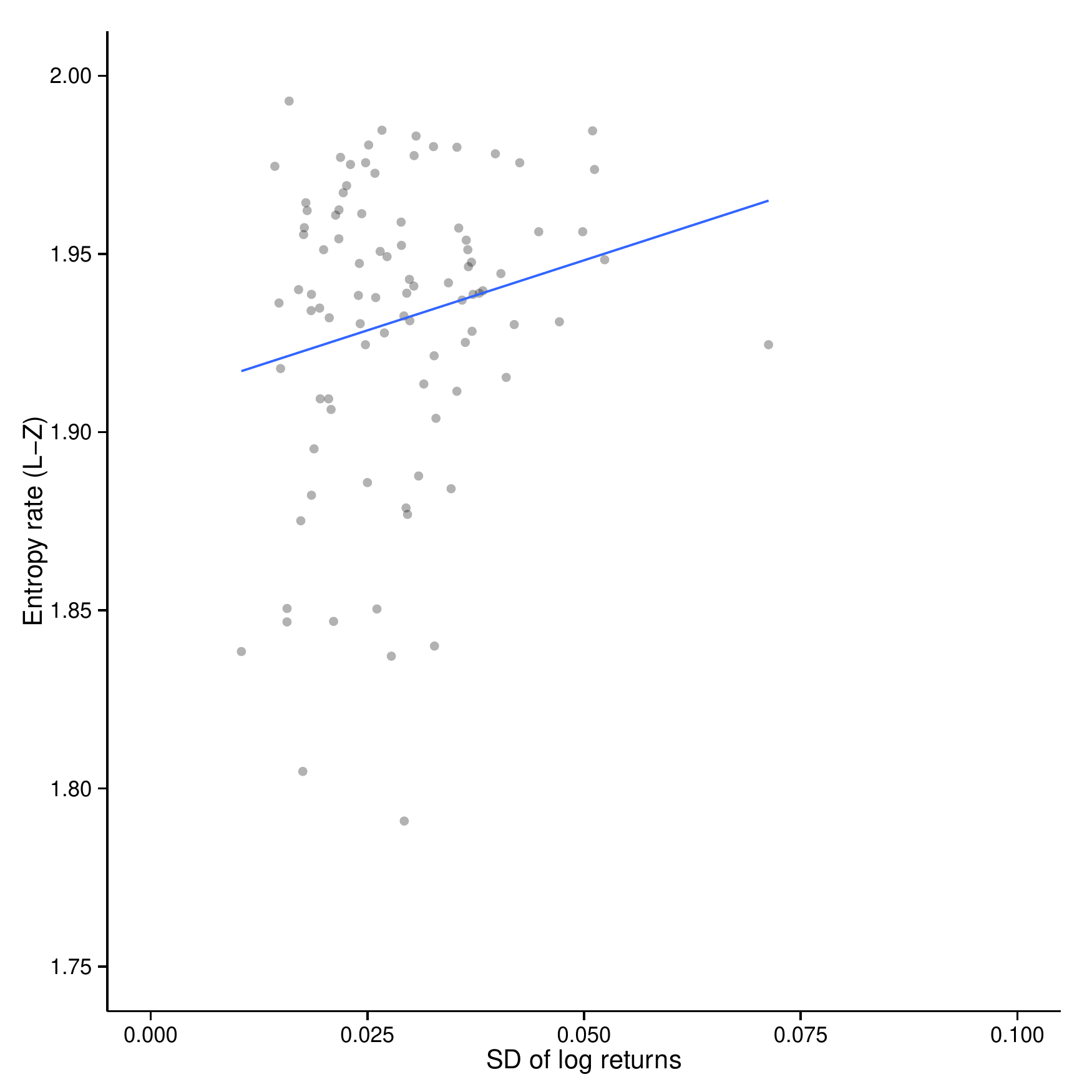}
\caption{{Entropy rate vs. SD of log returns (100 most influential)}}
\label{fig:lzstdev100}
\end{figure}

On Fig. \ref{fig:lzstdev} we can see the scatterplot showing the correlation between entropy rate and standard deviation of log returns for all the observed data series. As we can see the correlation is indeed negative. But on Fig. \ref{fig:lzstdev100} we can see the same data for only the 100 data points with the biggest Markov centrality measure. And this confirms our suspicion that for the most important companies the correlation is positive, that is the bigger the diversity in log returns the closer the market is to the efficient market hypothesis.

\section{Conclusions}

In this paper we have shown that the banking sector had too much influence on Warsaw's market in 2008-2010, therefore propagating the financial crisis all over the market, but that influence is now dwindling spelling doom for the financial crisis define as the negative influence of the banking sector started in 2007. Similar studies should be performed on other markets to confirm whether this is robust, as well as compare the speed of this recovery on different markets. Further studies should also be perfomed treating companies from the same sector in networks as clusters, and the results should be compared to the ones obtained in this study. We have also shown that while for the most influential companies the correlation between entropy rate and standard deviation of log returns is positive, the same is significant and negative for the whole market. That means when there is more variability then the prices are more predictable, which is somewhat counterintuitive and requires further studies to determine the nature and causes of this phenomenon.

\section*{Acknowledgements}

We would like to thank Thomas Kern for his help with reproducing horizon graphs with ggplot2.

\bibliographystyle{agsm}
\bibliography{prace}
\end{document}